\documentstyle [12pt] {article}
\begin {document}

\def\gsim{\:\raisebox{-0.5ex}{$\stackrel{\textstyle>}{\sim}$}\:}

\begin {flushright}
TIFR/TH/93-64 \\
December 1993
\end{flushright}

\bigskip

\begin {center}

{\Large {\bf Signatures of an Invisibly Decaying Higgs Particle at
LHC}} \\

\bigskip

{\large Debajyoti Choudhury$^a$ and D.P. Roy$^b$\footnote{E-mail :
DPROY@TIFRVAX.BITNET}} \\
\medskip
{\em $^a$ Max-Planck-Institut fur Physik, D-80805 Munchen, Germany} \\
\smallskip
{\em $^b$ Theory Group, Tata Institute of Fundamental Research, Bombay
400 005, India}
\end{center}

\begin {abstract}
The Higgs particle can decay dominantly into an invisible channel in
the Majoron models. We have explored the prospect of detecting such a
Higgs particle at LHC via its associated production with a gluon, Z or
W boson. While the signal/background ratio is too small for the first
process, the latter two provide viable signatures for detecting such a
Higgs particle.
\end {abstract}

\newpage

The signatures for Higgs particle detection at LEP and the future
hadron colliders (LHC/SSC) have been extensively studied in the
framework of the standard model (SM) and its supersymmetric (SUSY)
extensions [1,2]. There exist some extensions of the SM, however, with
a qualitatively different signature for the Higgs particle. These
extensions are generically called the Majoron models [3-7] and have
been quite popular e.g. in the context of generating neutrino
mass. They are characterized by the existence of a Goldstone boson
(the Majoron). Since the coupling of this Goldstone boson to the Higgs
particle is not required to be small on any theoretical or
phenomenological grounds, the Higgs particle could decay into an
invisible channel containing a Majoron pair [7-9]. Indeed the
importance of extending the Higgs search to this invisible decay
channel has been repeatedly emphasized over the past decade
[8,9]. However, quantitative investigations along this line have only
started very recently [10,11].

The key features shared by essentially all Majoron models is a
spontaneously broken global $U(1)$ symmetry and a complex $SU(2)
\times U(1)$ singlet scalar field $\eta$ transforming nontrivially
under the global $U(1)$. The spontaneous breaking of the global $U(1)$
generates a massless Goldstone boson, the Majoron $J \equiv Im
\eta/\sqrt 2$, and a massive scalar $\eta_R \equiv Re \eta/\sqrt
2$. The latter mixes with the massive neutral component $\phi_R$ of
the standard Higgs doublet through a quartic term $\phi^\dagger \phi
\eta^\dagger \eta$ in the scalar potential. Thus one has two massive
physical scalars $$ H = \cos \theta \phi_R + \sin \theta \eta_R ~,~~S
= \cos \theta \eta_R -
\sin \theta \phi_R ~,
\eqno (1)
$$
where the mixing angle can be chosen to lie in the range 0-45$^\circ$, so
that the $H$ and $S$ have dominant doublet and singlet components
respectively. The above quartic term also generates the following
couplings of $H$ and $S$ to the massless Goldstone boson $J$ :
$$
{\cal L} = {(\sqrt 2 G_F)^{1/2} \over 2} \tan \beta \left[M_S^2 \cos \theta S
J^2
- M_H^2 \sin \theta H J^2 \right]~,
\eqno (2)
$$ where $\tan \beta = <\phi>/<\eta>$ is the ratio of the two vacuum
expectation values [8,9]. The resulting decay widths of $H, S$ into
the invisible channel $(JJ)$ relative to the dominant SM channel $(b
\bar b)$ are
$$
\Gamma_{H \rightarrow JJ}/\Gamma_{H \rightarrow b \bar b} \simeq {1 \over
12} \left({M_H \over m_b}\right)^2 \tan^2 \theta \tan^2 \beta \left(1 -
{4m^2_b \over M_H^2} \right)^{-3/2}~,
\eqno (3)
$$
$$
\Gamma_{S \rightarrow JJ}/\Gamma_{S \rightarrow b \bar b} \simeq {1 \over
12} \left({M_S \over m_b}\right)^2 \cot^2 \theta \tan^2 \beta \left(1 -
{4m^2_b \over M_S^2} \right)^{-3/2}~.
\eqno (4)
$$ The large mass ratio $(M_{H,S}/m_b)^2$ on the rhs implies that the
invisible decay channel could dominate for $S$ as well as $H$ over a
large range of the parameters $\tan \theta$ and $\tan
\beta$.\footnote{Both the parameters depend on the scale of the global
$U(1)$ breaking relative to the $SU(2) \times U(1)$ breaking scale, on
which there are no severe phenomenological constraints.}

Although the eqs. (1-4) above were derived for the simplest model [3]
having 1 singlet and 1 doublet scalar fields, similar considerations
hold for those having a larger Higgs content [5-8] or a larger global
symmetry group than $U1$ [12]. It may be added here that, the Higgs
particles can also decay invisibly in the SUSY models via a pair of
lightest superparticles (LSP). For the minimal supersymmetric standard
model (MSSM) this invisible decay mode has been shown to dominate only
over a tiny range of parameters for the scalar Higgs particles but
over a larger range for the pseudoscalar [13].

Thus it is important to extend the Higgs search strategies at LEP and LHC
to cover the possibility of a dominantly invisible decay. This is simple at
LEP, since the dominant channel for Higgs search is the same for the SM
and the invisible decays -- i.e. the missing energy channel with one or
two jets [10,11]. It corresponds to the Bjorken production process
$$
e^+ e^- \buildrel Z \over \rightarrow Z^\ast H
\eqno (5)
$$
followed by $Z^\ast \rightarrow \nu \bar \nu ~, ~H \rightarrow b \bar b$
for the SM decay and $Z^\ast \rightarrow q \bar q ~,~ H \rightarrow JJ$
for the invisible decay. Indeed the larger branching fraction of $Z^\ast$
into quarks implies a larger event rate for the latter case. On the other
hand the production cross-section would be suppressed by a factor of
$\cos^2 \theta$ in this case. Combining the two effects leads to a $M_H$
bound in the Majoron models, which is within $\pm$ 6 GeV of the SM value
irrespective of the model parameters -- i.e. 60 $\pm$ 6 GeV [10,11]. A
similar correlation between the Higgs signatures for the two models is
expected to hold at LEP-II as well.

The present work is devoted to a systematic exploration of the
signatures for an invisibly decaying Higgs particle search at the
proposed large hadron collider (LHC). To start with one notes that the
missing energy is not a measurable quantity at a hadron collider due
to the large energy loss along the beam pipe. One has to consider
instead the missing transverse momentum $({p\!\!\! /}_T)$. Thus one
must look at one of the following associated production processes
which dominate Higgs production at large $p_T$ -- (i) H + jet, (ii) H
+ Z, (iii) H + W and (iv) H + t $\bar {\rm t}$. We shall present below
the results of our analysis for the first three processes. We shall
see that the associated production of H with jet has too large a
background to be a viable channel. On the other hand the associated
production of H with W or Z bosons are expected to give viable
signals.\footnote{These two channels have been considered before in
[14]. However, the results presented there have limited utility, since
they do not contain the ${p\!\!\! /}_T$ distributions nor the effects
of any ${p\!\!\! /}_T$ cuts on the signals and the corresponding
backgrounds.} We shall not discuss the associated production of H with
t$\bar{\rm t}$, since it has been analysed recently in [15]. We shall
only comment on the relative merit of this channel vis a vis H + W(Z).

\medskip

\noindent(i)~\underbar{H + jet Channel} : The dominant contribution to
this process comes from gluon-gluon fusion via the top quark loop
$$
gg \rightarrow gH~,
\eqno (6)
$$
followed by the invisible decay $H \rightarrow JJ$. The cross-section for
(6) was computed for SSC/LHC energies in [16] in the $m_t \rightarrow
\infty$ limit and in [17,18] for finite values of $m_t$. In the latter
case the formalism is quite involved, containing several dilogarithmic
functions. The cross-section presented below has been calculated for $m_t
= $160 GeV using the code of ref. [18]. The dominant background comes from
the tree-level process
$$
qg \rightarrow qZ
\eqno (7)
$$
followed by $Z \rightarrow \nu \bar \nu$. The contribution from $q \bar q
\rightarrow gZ$ is about an order of magnitude smaller.

Figure 1 shows the jet + ${p\!\!\! /}_T$ cross-section coming from the
signal process (6) for $M_H $ = 120 GeV along with the background (7)
at the LHC energy of 14 TeV and a rapidity cut of $|y_{jet}| < 5$.
The two sets of curves shown have been calculated using the structure
functions and the QCD coupling parameters of DFLM [19] and MRS
[20]. Although the quantitative estimate of the cross-sections seems
to depend appreciably on this choice, the background is seen to be at
least a factor of 40 larger than the signal. We have checked that
reducing the rapidity cut leads to only a marginal decrease of this
factor. Thus the H + jet channel does not provide a viable signature
for an invisibly decaying Higgs particle.

\medskip

\noindent (ii)~\underbar{H + Z Channel} : The dominant contribution to
this signal comes from the associated Bjorken process
$$
q \bar q \buildrel Z^\ast \over \rightarrow HZ
\eqno (8)
$$
followed by $H \rightarrow JJ$ and $Z \rightarrow \ell^+ \ell^-$,
where $\ell$ represents both electron and muon. The dominant
background comes from $$ q \bar q \rightarrow ZZ
\eqno (9)
$$ followed by $\nu \bar \nu$ decay of one $Z$ and $\ell^+ \ell^-$
decay of the other. There are also one-loop contributions to the
signal from $gg\rightarrow ZH$ (via top quark) and the background from
$gg \rightarrow ZZ$. The former is expected to add $\sim 10\%$ to the
signal [21] and the latter $\sim 30\%$ to the background [22] at the
LHC energy. Moreover the $p_T^Z$ dependence of the one-loop and the
corresponding tree-level cross-sections are very similar, particularly
for the background. Thus including these one-loop contributions would
only reduce the signal/background ratio by $\sim 20\%$. Therefore, it
is adequate for our purpose to restrict to the dominant tree-level
contributions.

We have calculated the dilepton + ${p\!\!\! /}_T$ cross-sections
coming from the signal (8) and the background (9) for the LHC energy
of = 14 TeV and the following acceptance cuts [23],
$$
|y_{{p\!\!\!/}_T}| < 5 ~,~ |y_\ell| < 3 ~,~ p^\ell_T > 20 ~{\rm GeV}~.
\eqno (10)
$$

Figure 2 shows the ${p\!\!\! /}_T$ distributions of the signal and
background cross-sections for $M_H$ = 120 and 160 GeV, assuming the
DFLM structure functions [19]. We have checked that using the MRS [20]
(EHLQ [24]) structure functions instead would raise (lower) these
cross-sections by only $\sim 10\%$. While the cross-sections are seen
to go down rapidly with increasing ${p\!\!\! /}_T$, there is a marked
enhancement of the signal/background ratio. In particular, for a
${p\!\!\! /}_T >$ 200 GeV cut, the signal becomes about half the size
of the background. Of course the reason for this is that, while the
signal (8) is a s-channel process, the background (9) has an
additional suppression factor coming from the t-channel propagator.

Table 1 shows the integrated signal and background cross-sections for
the missing-$p_T$ cuts of ${p\!\!\! /}_T >$ 100 and 200
GeV.\footnote{While the typical missing$- p_T$ cut assumed for LHC
studies is ${p\!\!\! /}_T >$ 200 GeV [23], we feel that the
accompanying dilepton pair should make it feasible to reduce this cut
to 100 GeV.} Since the size of the $ZZ$ background can be estimated
from the $\ell^+ \ell^- \ell^+ \ell^-$ channel, it should be possible
to extract the $ZH$ signal if
$$ S/\sqrt B \gsim 5~,
\eqno (11)
$$
where $S$ and $B$ denote the numbers of signal and background events.
Table 1 lists this ratio for both the low and high luminosity options of
LHC, corresponding to typical integrated luminosities of 10 and 100
events/fb respectively. One sees from this list that it should be possible
to search for an invisibly decaying Higgs particle upto $M_H$ = 120 GeV
(200 GeV) at the low (high) luminosity option of LHC. This assumes of
course a modest mixing angle $(\cos^2 \theta \simeq 1)$, so that one does
not pay an appreciable price at the production vertex.

\medskip

\noindent (iii)~\underbar{H + W Channel} : The signal comes from the process
$$
q \bar q^\prime \buildrel W^\ast \over \rightarrow WH~,
\eqno (12)
$$
followed by the decays $W \rightarrow \ell \nu$ and $H \rightarrow JJ$,
resulting in lepton + ${p\!\!\! /}_T$ events. The corresponding irreducible
background is from
$$
q \bar q^\prime \rightarrow W Z~,
\eqno (13)
$$
followed by $W \rightarrow \ell \nu$ and $Z \rightarrow \nu \nu$. There is
additional background from $W$ + multijet events, which can be
effectively suppressed however by demanding a transverse mass cut
$$
M_{\ell {p\!\!\! /}_T} = \left[(p^\ell_T + {p\!\!\! /}_T)^2 - (\vec p_T^\ell +
\vec {p\!\!\! /}_T)^2 \right]^{1/2} > 100 GeV.
\eqno (14)
$$
This constraint is automatically satisfied by the above signal (as well as
the irreducible background) process for ${p\!\!\! /}_T >$ 200 GeV.
Therefore, we shall concentrate on this region of missing-$p_T$. There
remains one potentially serious background here -- i.e. from
$$
t \bar t \rightarrow WWb \bar b \rightarrow \ell \nu \ell \nu b \bar b~,
\eqno (15)
$$
where $p_T$ of one of the leptons (including $\tau$) is $<$ 20 GeV,
so that it can not be identified. The size of this background is an
order of magnitude larger than the signal (12). Note, however, that
the signal (12) has no hadronic jet activity apart from those coming
from the initial state QCD radiation, which would be soft and/or
collinear with the beam.  Therefore, it should be possible to suppress
the above background (15) by a suitable $p_T$ and/or angular cut on
the accompanying hadronic $(b)$ jets.  But we have not been able to
pursue this quantitatively since we had no access to an initial state
QCD radiation program.

The lepton + ${p\!\!\! /}_T$ cross-sections coming from the signal
(12) and the irreducible background (13) have been calculated for the
LHC energy of 14 TeV and the cuts of eq. (10). These cross-sections
are again insensitive to the choice of structure functions. Figure 3
shows the ${p\!\!\! /}_T$ distributions of the signal and background
for $M_H$ = 120 and 160 GeV using the DFLM structure functions. There
is again a marked improvement in the signal/background ratio with
increasing ${p\!\!\! /}_T$.  Indeed in the region of our interest,
${p\!\!\! /}_T >$ 200 GeV, the signal is as large as the background.

Table 1 shows the integrated cross-sections for the signal and
background for ${p\!\!\! /}_T$ cuts of $>$ 100 and 200 GeV. But we
shall concentrate on the 200 GeV cut for this process for the reason
mentioned above. In this case the $S/\sqrt B$ seems to be viable for
the invisibly decaying Higgs particle search upto $M_H$ = 200 GeV even
at the low luminosity option of LHC. And at the high luminosity option
one can even afford to make generous allowance for a suppression
factor $(\cos^2 \theta)$ at the production vertex.\footnote{It may not
be necessary to extend the probe beyond $M_H$ = 200 GeV, since in this
case the $H \rightarrow WW, ZZ$ modes are expected to dominate over
the $H \rightarrow JJ$ (or a pair of LSP).} One should of course bear
in mind the above mentioned reducible background from (15).

\medskip

\noindent(iv)~\underbar{H + t $\bar {\rm t}$ Channel} : This channel has been
analysed recently in [15]. It is useful to compare the resulting
signal with those of the $ZH$ and $WH$ channels. The final state
consists of $\ell$, ${p \!\!\! /}_T$ and 4 jets from the leptonic
decay of one of the $t$ quarks and hadronic decay of the other while
$H$ decays invisibly. With a ${p\!\!\! /}_T >$ 200 GeV cut, the signal
is comparable to the background for $M_H$ = 140 GeV; but the signal
size is only $\sim$ 0.5 fb. Thus it can only be viable at the high
luminosity option of LHC [15]. Besides this signal is far more
demanding on detector performance since it requires $b$ identification
as well as reconstruction of $W$ and $t$ masses from hadronic
jets. However, as emphasized in [15], this is the only channel
available for detecting an invisibly decaying pseudoscalar Higgs boson
of the SUSY model, since it does not couple to $ZZ$ and $WW$ channels.

In summary, the Higgs boson can have dominantly invisible decay in the
Majoron models as well as some SUSY models. We have explored the prospects
of detecting such a Higgs particle at LHC in the H + jet, H + Z and H + W
channels. While the signal is expected to be overwhelmed by the background
in the first case, one expects viable signals in the H + Z and H + W
channels.

We are grateful to Nigel Glover for providing us with the code for
computing the $gH$ cross-section as well as his collaboration in some of
these computations. Indeed it is only by his insistence that his name
appears here instead of the first page of this paper. We are also grateful
to Rohini Godbole, Anjan Joshipura, Saurav Rindani and James Stirling for
a number of useful comments.

\newpage

\noindent\underbar{\bf References}

\begin {enumerate}

\item
See e.g. J.F. Gunion, H. Haber, G.L. Kane and S. Dawson, The Higgs
Hunter's Guide (Addison-Wesley, Reading, MA, 1990).

\item
ALEPH Collaboration : D. Decamp et al, Phys. Rep. {\bf 216} (1992) 253.

\item
Y. Chikashige, R.N. Mohapatra and R.D. Peccei, Phys. Lett. {\bf 98B}
(1980) 265.

\item
D.B. Reiss, Phys. Lett. {\bf 115B} (1982) 217; F. Wilczek, Phys. Rev.
Lett. {\bf 49} (1982) 1549.

\item
G. Gelmini and M. Roncadelli, Phys. Lett. {\bf B99} (1981) 411.

\item
A.S. Joshipura and S.D. Rindani, Phys. Rev. {\bf D46} (1992) 3000; \hfill\break
A.S. Joshipura and J.W.F. Valle, Nucl. Phys. {\bf B397} (1993) 105.

\item
RindaniJ.C. Ramao, F. de Campos and J.W.F. Valle, Phys. Lett. {\bf B292} (1992)
329.

\item
R.E. Schrock and M. Suzuki, Phys. Lett. {\bf B110} (1982) 250; \hfill\break
L.F. Li, Y. Liu and L. Wolfenstein, Phys. Lett. {\bf B159} (1985) 45;
\hfill\break
E.D. Carlson and L.B. Hall, Phys. Rev. {\bf D40} (1985) 3187; \hfill\break
G. Jungman and M.A. Luty, Nucl. Phys. {\bf B361} (1991) 24.

\item
A.S. Joshipura and S.D. Rindani, Phys. Rev. Lett. {\bf 69} (1992) 3269.

\item
B. Brahmachari, A.S. Joshipura, S.D. , D.P. Roy and K. Sridhar,
Phys. Rev. {\bf D48} (1993) 4224.

\item
ALEPH Collaboration : D. Buskulic et al, Phys. Lett. {\bf B313} (1993) 312.

\item
J.D. Bjorken, Invited Talk at the Symp. on SSC Laboratory, Corpus Christi,
Texas, October 1991, SLAC-PUB-5673(1991).

\item
A. Djouadi, J. Kalinowski and P.M. Zerwas, Z. Phys. {\bf C57} (1993) 569.

\item
S.G. Frederiksen, N.P. Johnson, G.L. Kane and J.H. Reid,
SSCL-preprint-577, July 1992 (Unpublished).  See also J.C. Romao, J.L.
Diaz-Cruz, F. Campos and J.W.F. Valle, FTUV/92-39, Nov. 1992
(Unpublished).

\item
J.F. Gunion, UCD-93-28, Revised version, December 1993.

\item
M. Chaichian, I. Liede, J. Lindfors and D.P. Roy, Phys. Lett. {\bf B198}
(1987) 416 and {\bf B205} (1988) 595E.

\item
R.K. Ellis, I. Hinchliffe, M. Soldate and J.J. Van der Bij, Nucl. Phys.
{\bf B297} (1988) 221.

\item
U. Baur and E.W.N. Glover, Nucl. Phys. {\bf B339} (1990) 38.

\item
M. Diemoz, F. Ferroni, E. Longo and G. Martinelli, Z. Phys. {\bf C39}
(1988) 21. We have used a parametrisation of these structure functions by
M.G. Gluck, R.M. Godbole and E. Reya, DO-TH-89/16.

\item
A.D. Martin, R.G. Roberts and W.J. Sterling, Phys. Rev. {\bf D47} (1993) 867.

\item
B.A. Kniehl, Phys. Rev. {\bf D42} (1990) 2253.

\item
E.W.N. Glover and J.J. Van der Bij, Nucl. Phys. {\bf B321} (1989) 561.

\item
C. Albajar et al, Proc. of ECFA-LHC Workshop, Vol. II, CERN90-10(1990) 621.

\item
E. Eichten, I. Hinchliffe, K. Lane and C. Quigg, Rev. Mod. Phys. {\bf 56}
(1984) 579.

\end {enumerate}

\newpage

\begin{center}
{\bf Table 1} \\
The integrated cross-sections for the $ZH$ and $WH$ \\
signals for $M_H = 120,160,200$ GeV along \\
with the corresponding backgrounds. \\
The signal/$\sqrt {\rm background}$ ratios \\
are also shown for the integrated luminosity of \\
10 (100) events/$fb$.
\end{center}

\medskip

\[
\begin{tabular} {|l|l|l||l|l|}
\hline
Process&\multicolumn{2}{|c||} {${p\!\!\!/}_T > 100$ {\rm GeV}}
       &\multicolumn{2}{|c|} {${p\!\!\!/}_T > 200$ {\rm GeV}}  \\
        \cline{2-5} & $\sigma(fb)$ & $S/\sqrt{B}$  &  $\sigma(fb)$
          & $S/\sqrt{B}$ \\ \hline
&&&& \\
$ZZ$ & 23.3 & & 3.5 & \\
&&&& \\
$ZH (120~ {\rm GeV})$ & 9.0 & 5.9 (18.6) & 1.9 & 3.2 (10.2) \\
&&&& \\
$ZH (160~ {\rm GeV})$ & 5.3 & 3.5 (10.9) & 1.4 & 2.4 (7.5) \\
&&&& \\
$ZH (200~ {\rm GeV})$ & 3.3 & 2.2 (6.8) & 1.1 & 1.9 (5.9) \\\hline
&&&& \\
$WZ$& 38.5 & &3.1 & \\
&&&& \\
$WH (120~ {\rm GeV})$ & 26.0 & &3.3 & 5.9 (18.6) \\
&&&& \\
$WH (160~ {\rm GeV})$ & 16.1 & & 2.7 & 4.8 (15.2) \\
&&&& \\
$WH (200~ {\rm GeV})$ & 11.3 & & 2.4 & 4.3 (13.5) \\
\hline
\end{tabular}
\]

\newpage

\noindent\underbar{\bf Figure Captions}

\begin {description}

\item{Fig. 1.} The Higgs signal and the $Z$ background cross-sections for
the jet + missing-$p_T$ channel at LHC. The two sets of curves
correspond to the DFLM [19] and MRS [20] structure functions.

\item{Fig. 2.} The $HZ$ signal (dotted and dashed lines) and the $ZZ$
background (solid line) cross-sections for the dilepton + missing-$p_T$
channel at LHC, calculated using the DFLM structure functions [19].

\item{Fig. 3.} The $HW$ signal (dotted and dashed lines) and the $WZ$
background (solid line) cross-sections for the lepton + missing-$p_T$
channel at LHC, calculated using the DFLM structure functions [19].

\end {description}

\end {document}